\documentclass[twocolumn,superscriptaddress,amsmath,ammsymb]{revtex4-1}
\usepackage{epsfig}
\usepackage{amsmath}
\usepackage{amssymb}
\usepackage{graphicx}
\usepackage{dcolumn}
\usepackage{bm}
\usepackage[usenames]{color}
\usepackage[breaklinks]{hyperref}
\usepackage{soul}
\usepackage{ulem}
\hypersetup{colorlinks=true,allcolors=blue}

\begin{document}
	
\title{Quantum theory of the magnetochiral anisotropy coefficient in ZrTe$_5$}
\author{Yi-Xiang Wang}
\email{wangyixiang@jiangnan.edu.cn}
\affiliation{School of Science, Jiangnan University, Wuxi 214122, China}
\affiliation{School of Physics and Electronics, Hunan University, Changsha 410082, China}

\author{Fuxiang Li}
\email{fuxiangli@hnu.edu.cn}
\affiliation{School of Physics and Electronics, Hunan University, Changsha 410082, China}

\date{\today}

\begin{abstract}
Recent experiments performed nonreciprocal magnetotransport studies in ZrTe$_5$ and obtained a giant magnetochiral anisotropy (MCA) coefficient $\gamma'$. 
The existing theoretical analysis was based on the semiclassical Boltzmann equation.    
In this paper, we develop a full quantum theory to calculate $\gamma'$ and further explore the underlying physics.  
We reveal that the $xz$-mirror symmetry breaking term also breaks the parity symmetry of the system and leads to mixed selection rules and nonvanishing second-order conductivity $\sigma_{xxx}$.  The calculations show that $\gamma'$ decreases with the magnetic field, survives only to the weak impurity scatterings, and exhibits a nonmonotonous dependence on the strength of the $xz$-mirror symmetry breaking.  Our paper can provide deeper insights into the intrinsic nonreciprocal magnetotransport phenomena in the topological semimetal material.  
\end{abstract} 

\maketitle

\section{Introduction}

Symmetries in the microscopic dynamics are vital in determining the macroscopic physical responses to the external excitations.  One example is that the quantum anomalous Hall effect can emerge in a system with broken time-reversal symmetry (TRS)~\cite{Haldane, C.Z.Chang2013, C.Z.Chang2023}, while the quantum spin Hall effect is protected by the TRS~\cite{M.Z.Hasan, X.L.Qi}.  Another example is that magnetochiral anisotropy (MCA) will occur in a system with broken inversion symmetry~\cite{Y.Tokura}.  MCA means that, under a magnetic field $B$, the resistance of a three-dimensional (3D) crystal sample is nonreciprocal when an electric current flows in the opposite directions.  In this case, the nonreciprocal resistance $R$ includes both the linear and quadratic terms, and is written as 
\begin{align}
R=\frac{V_x}{I_x}=R_0(1+\gamma BI_x), 
\label{Vx}
\end{align} 
where $I_x$ is the electric current that we assume to flow along the $x$ direction, $V_x$ is the voltage, $R_0$ is the reciprocal resistance, and the coefficient $\gamma$ characterizes the magnitude of MCA. 
Equation~(\ref{Vx}) is valid for both the inner-product-type~\cite{Rikken2001} and the vector-product-type~\cite{Rikken2005} nonreciprocal behaviors, for which the magnetic field and electric current are parallel $\boldsymbol B\parallel \boldsymbol I$ and perpendicular $\boldsymbol B\perp \boldsymbol I$, respectively.  

On the other hand, the current density $j_x$ can be written as a function of the electric field $E_x$, 
\begin{align}
j_x=j_x^{(1)}+j_x^{(2)}=\sigma_{xx} E_x+\sigma_{xxx} E_x^2, 
\label{Jx}
\end{align}
in which $\sigma_{xx}$ and $\sigma_{xxx}$ are the first- and second-order longitudinal conductivities, respectively.  If the current is expressed as $I_x=L_yL_zj_x$ and the voltage as $V_x=L_xE_x$, with $L_{x,y,z}$ denoting the size of the sample, we obtain $\gamma=\frac{\sigma_{xxx}}{L_yL_zB\sigma_{xx}^2}$.  
Since $\gamma$ depends on the size and shape of the 3D system, we instead use the cross-section-independent coefficient 
\begin{align}
\gamma'=L_yL_z\gamma=\frac{\sigma_{xxx}}{B\sigma_{xx}^2}, 
\label{gamma'}
\end{align}  
to characterize MCA~\cite{T.Ideue}. 

The common bulk nonreciprocal response in a 3D crystal originates from the relativistic effect and is quite small~\cite{Rikken2005}. 
The experiments in trigonal tellurium revealed that MCA had an inner-product type with the coefficient as large as $\gamma'\sim10^{-8}$ m$^2$T$^{-1}$A$^{-1}$~\cite{Rikken2019}, where the inversion symmetry was broken by its helical crystal structure. 
The recent magnetotransport experiments in 3D ZrTe$_5$ at low temperatures reported that MCA had a vector-product type, in which $\gamma'$ could reach the order of $10^{-7}$ m$^2$T$^{-1}$A$^{-1}$ and was gigantic~\cite{Y.Wang, N.Wang}. 

Theoretically, to explain the behavior of $\gamma'$, the calculations of the conductivities, $\sigma_{xx}$ and $\sigma_{xxx}$, are the central issues.  The existing theoretical analysis~\cite{Y.Wang} employed the semiclassical Boltzmann equation (SBE) and identified the microscopic mechanism as the Fermi surface deformation due to the magnetic-field-induced Zeeman effect~\cite{T.Ideue, P.He}.  As the SBE can treat the first- and second-order distribution functions well, it has been widely used in nonlinear conductivity calculations~\cite{T.Morimoto, Z.Z.Du, M.X.Deng, Y.Gao, D.Sinha, I.Sodemann, S.Das, D.Kaplan}.  The nonlinear Hall effect has excited many interests in spin-orbit-coupled semiconductors and topological materials, including the intrinsic nonlinear Hall effect~\cite{C.Wang, H.Liu, D.Kaplan} and the Berry-curvature-dipole-induced nonlinear Hall effect~\cite{I.Sodemann, Z.Z.Du, Q.Ma, K.Kang, H.Watanabe2020, Y.Michishita}, in which the former is time-reversal odd and the latter is time-reversal even.  
However, in all the above studies~\cite{T.Morimoto, Z.Z.Du, M.X.Deng, Y.Gao, D.Sinha, I.Sodemann, S.Das, D.Kaplan, Q.Ma, K.Kang, C.Wang, H.Liu, H.Watanabe2020, Y.Michishita}, there have been no discussions about the nonlinear longitudinal magnetotransport in Dirac semimetals on the basis of Landau levels (LLs), which motivated this paper.  

In this paper, to calculate $\gamma'$ and further explore the underlying physics of the nonreciprocal magnetotransport in ZrTe$_5$~\cite{Y.Wang}, we develop a full quantum theory by using the density matrix method.  Note that the density matrix method was adopted to explore the second-order Hall conductivity in the magnetic multipole system~\cite{H.Watanabe2020} as well as the second-order photocurrent in topological antiferromagnets~\cite{H.Watanabe2021}.  
Such a quantized method has the advantages that it can deal well with the band characters of the system, and some other factors, such as the impurity scatterings. 
Combining the density matrix method with thermodynamical averaging over the current density operators, we derive the conductivity formula of $\sigma_{xx}$ and $\sigma_{xxx}$.  Then within the topological semimetal model describing 3D ZrTe$_5$~\cite{H.Weng, R.Y.Chen, Z.G.Chen, F.Tang, S.Galeski2021}, where the inversion symmetry is broken intrinsically, we make a systematic study of $\gamma'$ and analyze the effects of the magnetic field, the impurity scatterings, and the $xz$-mirror symmetry-breaking term. 

Our main results are given as follows: 
(i) the intrinsic $xz$-mirror symmetry-breaking term also breaks the parity symmetry of the system, which results in the mixed selection rules $n\rightarrow n,n\pm1,n\pm2,\cdots$, with $n$ denoting the LL index.  The mixed selection rules will in turn play a decisive role in determining the nonvanishing $\sigma_{xxx}$; 
(ii) $\gamma'$ decreases with the magnetic field $B$, and is nonvanishing at a weak $B$, which is consistent with the experimental observations~\cite{Y.Wang, N.Wang}; 
(iii) $\gamma'$ is susceptible to the impurity scatterings, and survives only with the weak impurity scatterings; 
and (iv) when the strength of the $xz$-mirror symmetry breaking increases,  $\gamma'$ grows first and then decreases. 
It is worth noting that although $\gamma'$ is consistent with the SBE results in the order of magnitude, the third point is distinct~\cite{Y.Wang} and the implications will be discussed.  
Our paper can provide deeper insights in understanding the MCA phenomenon in ZrTe$_5$ experiments.

\section{The first- and second-order conductivities} 

When an electric field ${\boldsymbol E}=E{\boldsymbol e}_x$ acts on a system, the electrostatic force will drive the electrons to deviate from its equilibrium state.  If we consider the electric field as a perturbation of the system, the nonequilibrium density matrix $\rho(t)$, up to the second order, is written as~\cite{L.Smrcka, G.D.Mahan} 
\begin{align}
\rho(t)=\rho_0+\rho_1e^{st}+\rho_2e^{2st}+\cdots.  
\end{align}
Here, $\rho_0=1/[e^{\beta(\hat H_0-\mu)}+1]$ is the equilibrium Fermi-Dirac distribution function at chemical potential $\mu$ and inverse temperature $\beta=\frac{1}{k_BT}$, with $k_B$ being the Boltzmann constant and $T$ the temperature.  $\rho_1$ and $\rho_2$ are the first- and second-order density matrices, respectively.  The factors $e^{st}$ and $e^{2st}$ are added to $\rho_1$ and $\rho_2$, with the parameter $s=0^+$ to ensure that they vanish at $t\rightarrow-\infty$.  

The evolution of the density matrix is determined by the quantum Liouville equation, which is written as $(\hbar=1)$~\cite{L.Smrcka},  
\begin{align}
i\frac{d\rho(t)}{dt}
=[\hat H_0+\hat H'e^{st},\rho(t)].  
\end{align}
Here $\hat H_0$ is the Hamiltonian of the system, and $\hat H'=-e{\boldsymbol E}\cdot{\hat{\boldsymbol r}}$ describes the interaction between the electric field and the system, with $\hat{\boldsymbol r}$ being the position operator.  After a direct calculation, $\rho_1$ and $\rho_2$ are obtained as (see Sec. I in the Supplemental Material (SM)~\cite{SuppMat})
\begin{align}
&\rho_1=\frac{1}{i}\int_0^\infty dt  e^{-st} [\hat H'(-t),\rho_0], 
\\
&\rho_2=\frac{1}{i^2}
\int_0^\infty dt e^{-2st}\int_0^\infty dt'
e^{-st'}\big[\hat H'(-t),[\hat H'(-t'),\rho_0]\big]. 
\end{align}
where $\hat H'(t)=e^{i\hat H_0t}\hat H'e^{-i\hat H_0t}$.  

The measurable first- and second-order current densities are calculated through the thermodynamical averaging over the current density operator $\hat{\boldsymbol j}=e\hat{\boldsymbol v}$, i.e., ${\boldsymbol j}^{(1)}=\text{Tr}[\rho_1 \hat{\boldsymbol j}]$ and ${\boldsymbol j}^{(2)}=\text{Tr}[\rho_2 \hat{\boldsymbol j}]$.  With the help of the Green's function, the first-order and second-order longitudinal conductivities, $\sigma_{xx}$ and $\sigma_{xxx}$, can be derived (see Secs. II and III in the SM~\cite{SuppMat}).  At zero temperature, their expressions read   
\begin{align}
&\sigma_{xx}=\frac{e^2\eta^2}{\pi V} \sum_{\boldsymbol k} \sum_n\sum_{n'}
\frac{|\langle \psi_n|\hat v_x|\psi_{n'}\rangle|^2}
{[(\mu-\varepsilon_n)^2+\eta^2][(\mu-\varepsilon_{n'})^2+\eta^2]}, 
\label{sigmaxx}
\end{align}
and
\begin{align}
\sigma_{xxx}=&\frac{2e^3}{V}\sum_{\boldsymbol k} 
\sum_{\varepsilon_n<\mu}\sum_{\varepsilon_{n'}>\mu}\sum_{n''}
\nonumber\\
&\text{Re}\big[
\frac{\langle \psi_n|\hat v_x|\psi_{n'}\rangle\langle \psi_{n'}|\hat v_x|\psi_{n''}\rangle\langle \psi_{n''}|\hat v_x|\psi_n\rangle}
{(\varepsilon_n-\varepsilon_{n'}-i\eta)^2 (\varepsilon_n-\varepsilon_{n''}+i\eta)^2}
\nonumber\\
&-\frac{\langle \psi_n|\hat v_x|\psi_{n''}\rangle\langle \psi_{n''}|\hat v_x|\psi_{n'}\rangle\langle \psi_{n'}|\hat v_x|\psi_n\rangle}
{(\varepsilon_n-\varepsilon_{n'}+i\eta)^2 (\varepsilon_{n''}-\varepsilon_{n'}-i\eta)^2}\big], 
\label{sigmaxxx}
\end{align}
respectively.  Here $V$ is the volume of the system, $\varepsilon_n$ is the energy and $|\psi_n\rangle$ the corresponding eigenstate, $\hat v_x$ is the velocity operator, and $\eta$ denotes the linewidth broadening that is introduced to represent the impurity scatterings phenomenologically.  For simplicity, we take $\eta$ as a constant for all eigenstates of the system. 

Note that the expression of $\sigma_{xx}$ is the same as those obtained from the well-known Kubo-Bastin formula~\cite{A.Bastin, H.W.Wang, Y.X.Wang2020}.  In $\sigma_{xxx}$, $|\psi_n\rangle$ and $|\psi_{n'}\rangle$ represent the initial and final state, respectively, while $|\psi_{n''}\rangle$ denotes the intermediate state.  Intuitively, since $\sigma_{xxx}$ does not include the $\hat v_y$ term, it has no relation to the Berry curvature and thus is distinct from the nonlinear Hall conductivity.  Equations~(\ref{sigmaxx}) and~(\ref{sigmaxxx}) relate the conductivities to the band structures of the system directly, which are transparent to interpretation.  

The presence of a magnetic field will drive the formation of the LLs.  In Eqs.~(\ref{sigmaxx}) and~(\ref{sigmaxxx}), the nonvanishing velocity operator matrix element determines the selection rules that will play important roles in the transport properties.  In $\sigma_{xx}$, there are no restrictions on the selection rules; they may take the conventional ones $n\rightarrow n,n\pm1$~\cite{Y.Jiang, Y.X.Wang2023, Z.Cai}, with $n$ the LL index, or the unconventional ones $n\rightarrow n\pm2$~\cite{Y.X.Wang2022}.  By comparison, in $\sigma_{xxx}$, there exist multiplications of three matrix elements that form a closed loop, requiring that the selection rules would not simply take $n\rightarrow n\pm1, n\pm2$; they should take $n\rightarrow n$ or other types.

\section{Topological Semimetal Model and LLs}

\begin{figure}
	\includegraphics[width=9cm]{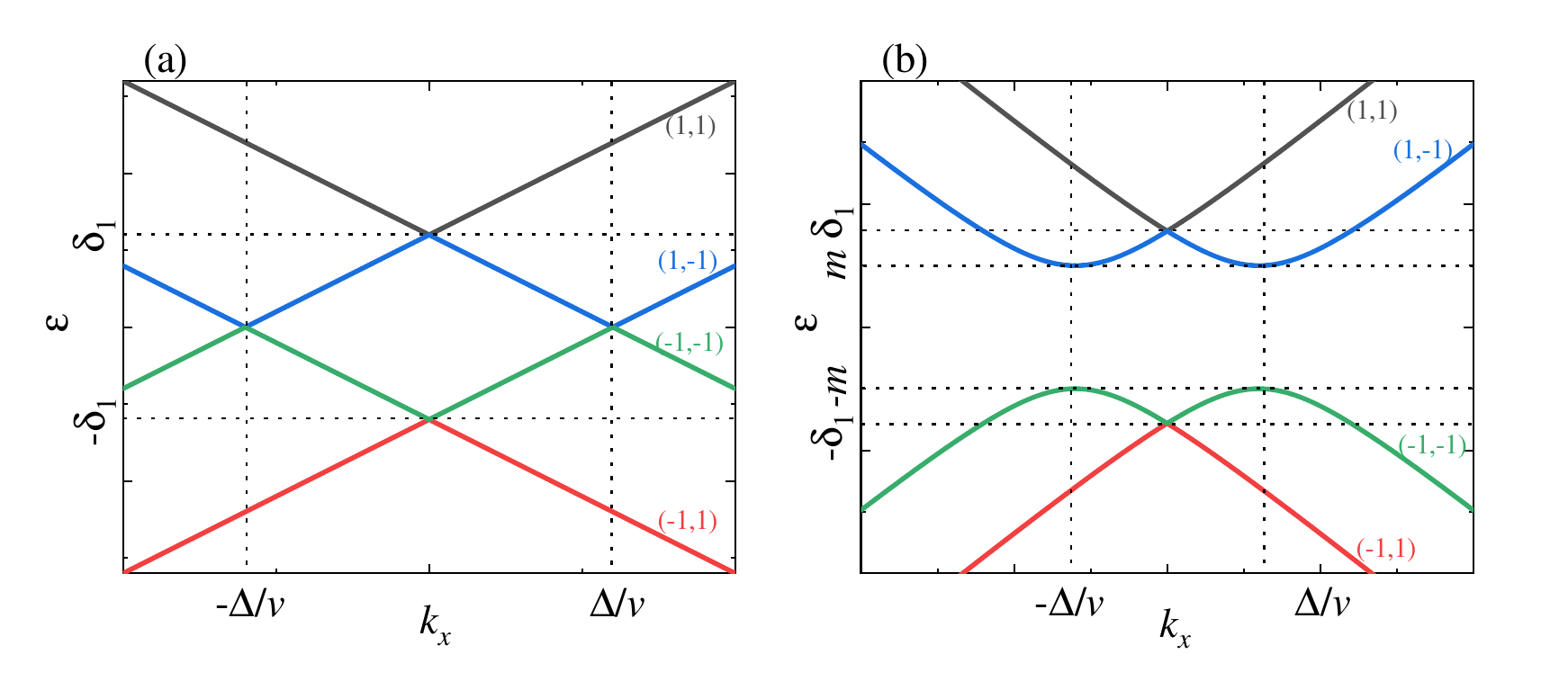}
	\caption{(Color online) The energy bands versus $k_x$ for $k_y=k_z=0$, with the band index $(s\lambda)$ being labeled.  In (a), when $m=0$, the system lies in the nodal line semimetal phase, and in (b), when $m\neq0$, the system is a gapped insulator.  We set the parameter $\xi=0$. }  
	\label{Fig1}
\end{figure}

We use the topological semimetal model to describe the low-energy excitations in 3D ZrTe$_5$.  In the four-component basis $\big(|+\uparrow\rangle,|-\uparrow\rangle,|+\downarrow\rangle,|-\downarrow\rangle\big)^T$, the Hamiltonian $\hat H_0(\boldsymbol k)$ is written as~\cite{H.Weng, R.Y.Chen, Z.G.Chen, F.Tang, S.Galeski2021} 
\begin{align}
\hat H_0(\boldsymbol k)=&v(k_x\sigma_z\otimes\tau_x+k_yI\otimes\tau_y)+v_zk_z\sigma_x\otimes\tau_x
\nonumber\\
&+m I\otimes\tau_z, 
\label{H0}
\end{align}
where $\sigma$ and $\tau$ are the Pauli matrices acting on the spin and orbit degrees of freedom, respectively.  $v$ is the Fermi velocity in the $xy$ plane, $v_z$ is the Fermi velocity in the $z$ direction, and $m$ denotes the Dirac mass.  Since it is generally accepted that ZrTe$_5$ is a narrow-gapped topological insulator, and the Fermi velocities satisfy $v\gg v_z$, we will take the model parameters $(v,v_z)=(6,0.5)\times10^5$ m/s, and $m=5$ meV, as extracted from the LL transition energies in the optical conductivity experiments of ZrTe$_5$~\cite{E.Martino, Y.Jiang}. 

In the 3D system, there exist the $xz$-mirror symmetry, $\hat{\cal M}_{xz}^{-1} {\hat H}_0(k_x,k_y,k_z)\hat{\cal M}_{xz}={\hat H}_0(k_x,-k_y,k_z)$, with the operator $\hat{\cal M}_{xz}=i\sigma_y\otimes\tau_z$,
the $xy$-mirror symmetry, $\hat{\cal M}_{xy}^{-1} {\hat H}_0(k_x,k_y,k_z)\hat{\cal M}_{xy}={\hat H}_0(k_x,k_y,-k_z)$, with $\hat{\cal M}_{xy}=\sigma_z\otimes I$, 
and the inversion symmetry $\hat{\cal I}^{-1}{\hat H}_0(\boldsymbol k)\hat{\cal I}={\hat H}_0(-\boldsymbol k)$, with $\hat{\cal I}=I\otimes\tau_z$.  
In a recent magnetotransport experiment~\cite{Y.Wang}, the inversion symmetry of 3D ZrTe$_5$ was demonstrated to be broken intrinsically, which was attributed to the staggered displacement of the Te atom along the $y$ direction.  Such inversion symmetry breaking terms are represented by
\begin{align}
\hat H_{IB}=\Delta I\otimes\tau_x+\xi\sigma_x\otimes\tau_y.   
\label{HIB}
\end{align}
in which the parameters $\Delta$ and $\xi$ denote the strengths of the $xz$- and $xy$-mirror symmetry breaking, respectively.  Clearly, the introduction of $\hat H_{IB}$ breaks the inversion symmetry as 
$\hat{\cal I}^{-1}{\hat H}_{IB}\hat{\cal I}=-{\hat H}_{IB}$.  

The energies of $\hat H_0+\hat H_{IB}$ are obtained as 
\begin{align}
\varepsilon_{s\lambda}(\boldsymbol k)=&s\Big[\big(\sqrt{v^2k_x^2+v_z^2k_z'^2}
+\lambda\sqrt{\Delta^2+\xi^2}\big)^2+v^2k_y'^2
\nonumber\\ 
&+m^2\Big]^\frac{1}{2}, 
\end{align}
where the index $s=\pm1$ denotes the conduction/valence band, and $\lambda=\pm1$ the two branches.  The redefined wave vectors are $vk_y'=vk_y\text{cos}\phi-v_zk_z\text{sin}\phi$, $v_zk_z'=vk_y\text{sin}\phi+v_zk_z\text{cos}\phi$, and tan$\phi=\frac{\xi}{\Delta}$, which is equivalent to rotating the $yz$ plane around the $x$ axis by angle $-\phi$.  In Fig.~\ref{Fig1}, the energy bands are plotted for $k_y=k_z=0$.  We see that the two $\lambda=1$ branches have a gap of $2\delta_1=2\sqrt{m^2+\Delta^2+\xi^2}$; another two $\lambda=-1$ branches cross at zero energy for $m=0$ [Fig.~\ref{Fig1}(a)] and are gapped for a nonvanishing $m$ [Fig.~\ref{Fig1}(b)].  The former occurs when  $k_x=\pm(\Delta^2+\xi^2-v_z^2k_z'^2)^\frac{1}{2}/v$ and $k_y=v_zk_z\text{tan}\phi/v$, indicating that the system lies in the nodal line semimetal phase and the Fermi surface has a torus shape~\cite{Y.Wang, D.Sinha}; while the latter means that the system is a gapped insulator. 

The one-dimensional Landau bands develop when a magnetic field is applied along the $z$ direction, $\boldsymbol B=B \boldsymbol e_z$.  To solve the LLs, we choose the Landau gauge $\boldsymbol A=-By\boldsymbol e_x$, which is minimally coupled to the crystal momentum through the Periels substitution $\boldsymbol p\rightarrow\boldsymbol p+e\boldsymbol A$.  Then we replace the momentum operators in $\hat H_0$ by the standard ladder operators $k_+\rightarrow \frac{\sqrt2}{l_B}\hat a^\dagger$ and $k_-\rightarrow\frac{\sqrt2}{l_B}\hat a$, with the magnetic length $l_B=\frac{1}{\sqrt{eB}}$.  The total Hamiltonian $\hat H_0+\hat H_{IB}$ under a magnetic field is written as 
\begin{align} 
\hat H_B&=\begin{pmatrix}
m& P\hat a+\Delta& 0& v_zk_z-i\xi
\\
P\hat a^\dagger+\Delta& -m& v_zk_z+i\xi& 0
\\
0& v_zk_z-i\xi& m& -P\hat a^\dagger+\Delta  
\\
v_zk_z+i\xi& 0& -P\hat a+\Delta& -m
\end{pmatrix}, 
\end{align}
in which the parameter $P=\frac{\sqrt2}{l_B}v$. 

\begin{figure}
	\includegraphics[width=9.2cm]{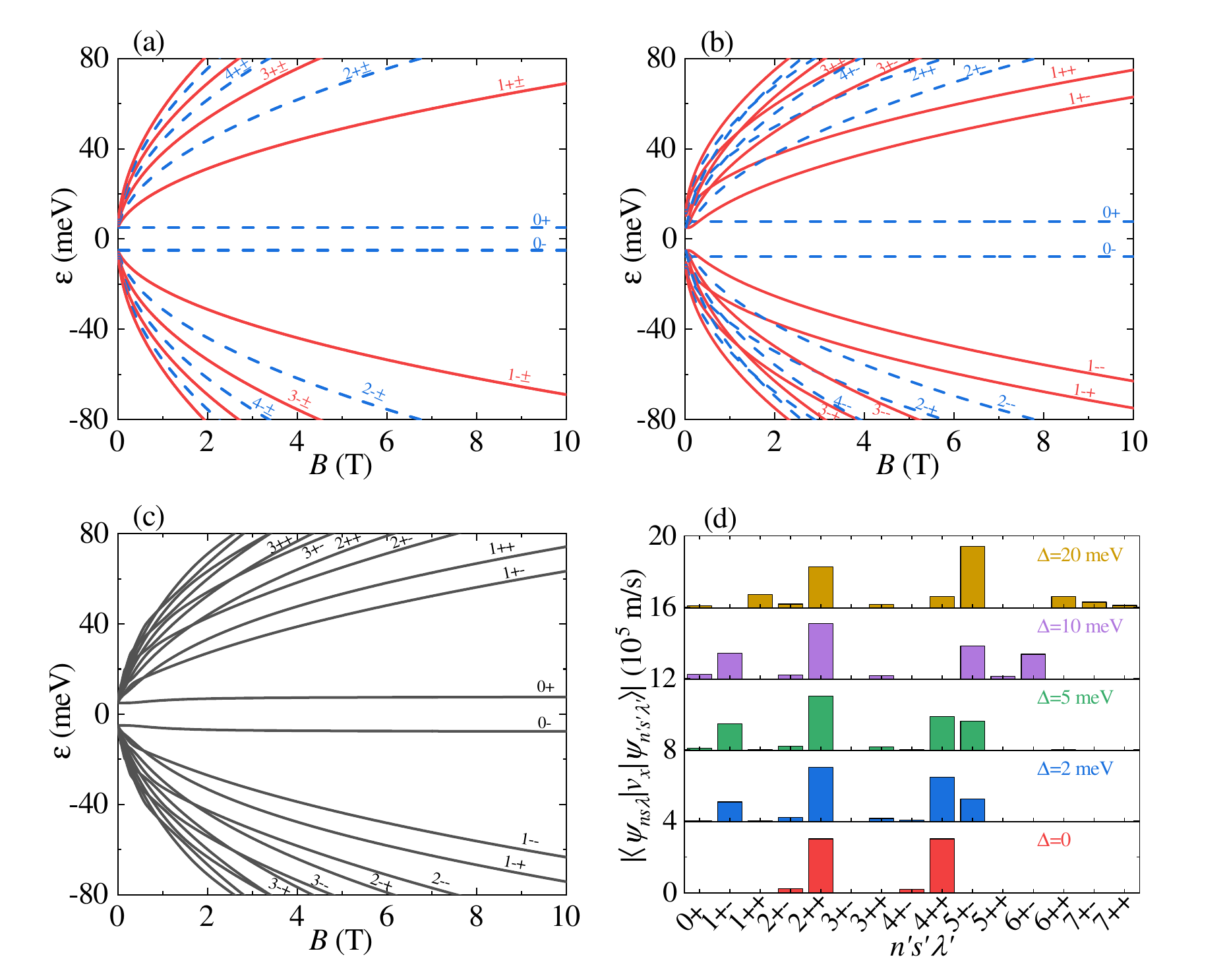}
	\caption{(Color online) The LL spectra in (a)$-$(c), with the index $ns\lambda$ being labeled, and the velocity operator matrix element $|\langle\psi_{ns\lambda}|\hat v_x|\psi_{n's'\lambda'}\rangle|$ between the initial state  and the final state for different $\Delta$ in (d), with the initial state index chosen as $(3--)$.  The parameters are set as $(\Delta,\xi)=(0,0)$ in (a), $(\Delta,\xi)=(0,6)$ meV in (b), $(\Delta,\xi)=(10,6)$ meV in (c), and the magnetic field $B=1$ T and $\xi=6$ meV in (d).  In (a) and (b), each LL owns a definite even or odd parity that are represented by the red solid or blue dashed lines, respectively.  In (d), the neighboring bars are shifted vertically for clarity. }  
	\label{Fig2}
\end{figure}

When $\Delta=0$, the LLs can be solved analytically.  For the $n\geq1$ LLs, we use the trial wave function $\psi_n=\big(c_n^1|n-1\rangle, c_n^2|n\rangle, c_n^3|n\rangle, c_n^4|n-1\rangle\big)^T$, with $|n\rangle$ being the harmonic oscillator state defined as $\hat a^\dagger \hat a|n\rangle=n|n\rangle$.  The LL energies are obtained as  
\begin{align}
\varepsilon_{ns\lambda}(k_z)=s\Big[v_z^2k_z^2+m^2+\big(P\sqrt n+\lambda\xi\big)^2\Big]^\frac{1}{2}. 
\label{LLn}
\end{align}
For the zeroth LLs, we use the trial wave function $\psi_0=\big(0,c_0^2|0\rangle, c_0^3|0\rangle,0\big)^T$. The energies are given as
\begin{align}
\varepsilon_{0s}(k_z)=s\big(v_z^2k_z^2+m^2+\xi^2\big)^\frac{1}{2}, 
\label{LL0}
\end{align}
which remain unchanged to the magnetic field.

The LL spectra are displayed in Figs.~\ref{Fig2}(a)$-$\ref{Fig2}(c).  When $\xi=0$, the $\lambda=\pm$ branches are degenerate [Fig.~\ref{Fig2}(a)]. 
The nonvanishing $\xi$ will break the twofold degeneracy of the $n\geq1$ LLs [Fig.~\ref{Fig2}(b)].  
At weak magnetic fields, when $\varepsilon_{n'++}>\varepsilon_{n+-}$ with $n'<n$, the LL crossings at higher energies are clearly seen, leading to the separation of the $ns\pm$ LLs. 
In the quantum Hall measurements, the LL crossings were also caused by broken inversion symmetry, which can induce complicated filling factors with the varying carrier density as well as the magnetic field~\cite{A.C.Lygo, D.A.Kealhofer}.  At strong magnetic fields with $P\sqrt n\gg\xi$, the $ns\pm$ LLs will rearrange and pair with each other.

\section{Selection Rules}

\begin{figure*}
	\centering
	\includegraphics[width=18cm]{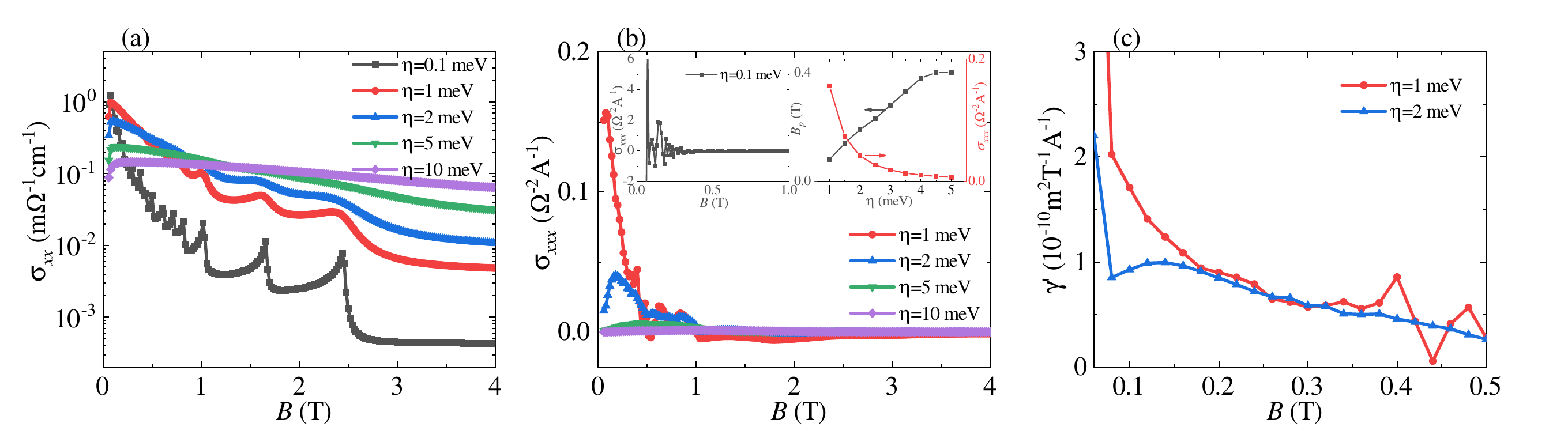}
	\caption{(Color online) (a)$-$(b) The conductivities $\sigma_{xx}$ and $\sigma_{xxx}$ versus the magnetic field $B$ for different linewidths $\eta$.  In (b), the left inset shows that, with $\eta=0.1$ meV, $\sigma_{xxx}$ exhibits strong oscillations, and the right inset shows the magnetic field $B_p$ and the corresponding conductivity $\sigma_{xxx}(B_p)$ versus $\eta$.  (c) The MCA coefficient $\gamma'$ at $\eta=1$ and $2$ meV.  We take the parameters $(\Delta,\xi)=(10,6)$ meV, and the chemical potential $\mu=30$ meV. }  
	\label{Fig3}
\end{figure*}

When $\Delta=0$, the Hamiltonian $H_B$ owns the parity symmetry with the operator $\hat{\cal P}=(-1)^{\hat a^\dagger \hat a}\sigma_z\otimes\tau_z$~\cite{T.Devakul, Y.X.Wang2022}.  This means that each LL carries a definite even or odd parity, as represented by the red solid or blue dashed lines in Figs.~\ref{Fig2}(a) and~\ref{Fig2}(b).  For the velocity operator $\hat v_x=v\sigma_z\otimes\tau_x$, because $\hat{\cal P}^{-1}\hat v_x\hat{\cal P}=-\hat v_x$, it has an odd parity.  Thus, the nonvanishing velocity operator matrix element $\langle \psi_n|\hat v_x|\psi_{n'}\rangle$ requires that the initial and final states own opposite parities, leading to the conventional selection rule $n\rightarrow n\pm1$~\cite{Y.Jiang, Y.X.Wang2023, Z.Cai}. 

When $\Delta$ is finite, $\hat H_B$ needs to be solved with numerics: in the infinite Hilbert space  spanned by the harmonic oscillator state $|n\rangle$, the LLs are obtained by diagonalizing $H_B$ after truncating the Hilbert space at the cutoff $N_c$.  In the calculations, we set $N_c=200$ to obtain the well-convergent results.  Now since $\hat{\cal P}^{-1}(\Delta\tau_x)\hat{\cal P}=-\Delta\tau_x$, the parity symmetry of the system is broken and the LLs will not carry a definite parity.  With a finite $\Delta=10$ meV, the LL crossings remain at higher energies and their movements are clearly seen [Fig.~\ref{Fig2}(c)].  Note that as $\hat{\cal P}^{-1}(\xi\sigma_x\otimes\tau_y)\hat{\cal P}=\xi\sigma_x\otimes\tau_y$, a finite $\xi$ does not break the parity symmetry of the system.  

To find out the selection rules, in Fig.~\ref{Fig2}(d), we display the velocity operator matrix element $|\langle \psi_{ns\lambda}|\hat v_x|\psi_{n's'\lambda'}\rangle|$ between the initial state $|\psi_{ns\lambda}\rangle$ and the final state $|\psi_{n's'\lambda'}\rangle$ for different $\Delta$.  The magnetic field is set as $B=1$ T and the initial state is chosen as $(3--)$. 
We see that when $\Delta=0$, the matrix elements are nonvanishing only for the final states $(2+\pm)$ and $(4+\pm)$, from which the selection rules are inferred as the conventional ones $n\rightarrow n\pm1$~\cite{Y.Jiang, Y.X.Wang2023, Z.Cai} and agree with the above parity symmetry analysis.  Actually, in such an analytically solvable system, the velocity operator cannot generate the transitions between the states with large $n$ difference.  As a result, the selection rules $n\rightarrow n\pm(2m+1)$, with $m\geq1$, are prohibited, although they meet the requirements of parity symmetry.  

When $\Delta=2$ meV and the parity of each LL is broken, we find that, in addition to $n\rightarrow n\pm1$, the selection rules can take $n\rightarrow n$ as well as $n\rightarrow n\pm2$.  
Further increasing $\Delta$, the LL eigenstate $|\psi_n\rangle$ may involve the harmonic oscillator state $|n'\rangle$ that is far away from $|n\rangle$, leading to more selection rules.  When $\Delta=20$ meV, the selection rules can even take $n\rightarrow n+4$.  Therefore, we suggest that the broken parity symmetry by the $\Delta$ term can lead to the mixed selection rules, $n\rightarrow n,n\pm1,n\pm2,\cdots$, which makes $\sigma_{xxx}$ nonvanishing.

\section{MCA coefficient $\gamma'$}

To obtain the MCA coefficient $\gamma'$, we need to calculate the conductivities $\sigma_{xx}$ and $\sigma_{xxx}$ by using Eqs.~(\ref{sigmaxx}) and~(\ref{sigmaxxx}).  The results are plotted in Figs.~\ref{Fig3}(a) and~\ref{Fig3}(b) as functions of the magnetic field $B$ for different linewidths $\eta$.  We fix the chemical potential at $\mu=30$ meV to ensure a low carrier density in the 3D system~\cite{F.Tang, E.Martino}.  

Firstly, we study $\sigma_{xx}$.  
In the limiting clean case of $\eta=0.1$ meV, $\sigma_{xx}$ exhibits a series of oscillation peaks that sit on a descending background.  
This is because under a weak magnetic field, the LLs are densely distributed.  
With increasing $B$, the LLs will cross the chemical potential $\mu$ one by one, resulting in the oscillation peaks of $\sigma_{xx}$~\cite{D.Shoenberg, Y.X.Wang2023}.  
Meanwhile, the decreasing LL number around $\mu$ leads to the descending background.  
When $B>2.44$ T, the system enters the extreme quantum limit, with all electrons confined to the zeroth LL~\cite{Z.Cai, S.Galeski2022, W.Wu}.  When  $\eta$ grows, in the quantum limit, $\sigma_{xx}$ increases, whereas in the quantum oscillation regime, $\sigma_{xx}$ exhibits a nonmonotonous variation, which is consistent with the previous studies~\cite{Y.X.Wang2020, Y.X.Wang2023}. 

Next, we study $\sigma_{xxx}$.  
When $\eta=0.1$ meV, $\sigma_{xxx}$ exhibits strong oscillations [the left inset of Fig.~\ref{Fig3}(b)], but will quickly drop to zero at a relatively large magnetic field.  
The former behavior can also be explained by the densely distributed LLs, and the latter is attributed to the enlarged LL spacings.  
With increasing impurity scatterings, the neighboring oscillations will merge together.
For weak impurity scatterings of $\eta=1$ and $2$ meV, we see that $\sigma_{xxx}$ first increases with $B$; after reaching its peak value at the magnetic field $B_p$, $\sigma_{xxx}$ then decreases to zero.  
To see this behavior more deeply, in the right inset of Fig.~\ref{Fig3}(b), we plot $B_p$ and the corresponding conductivity $\sigma_{xxx}(B_p)$ as functions of $\eta$.  We observe that $B_p$ increases with $\eta$ and is saturated when $\eta\geq4$ meV; while $\sigma_{xxx}$ decreases steadily to zero.  
Such a dependence of $\sigma_{xxx}$ on the magnetic field is consistent with the second-order longitudinal resistance observations in the experiment~\cite{N.Wang}, where temperature plays a similar role as the impurity scatterings here.  Further increasing $\eta$ to $\eta\geq5$ meV, $\sigma_{xxx}$ is completely suppressed.  These results indicate that $\sigma_{xxx}$ is evident only at a weak magnetic field, and is susceptible to the impurity scatterings.  

After obtaining $\sigma_{xx}$ and $\sigma_{xxx}$, we calculate the MCA coefficient $\gamma'$ by using Eq.~(\ref{gamma'}).  According to the above analysis, we see that $\gamma'$ depends on the impurity scatterings, which is different from the SBE.  In the SBE, the relaxation time $\tau$ enters through the time evolution of the distribution function, based on which $\sigma_{xx}\sim\tau$ and $\sigma_{xxx}\sim\tau^2$ are obtained, resulting in the independence of $\gamma'$ on the impurity scatterings~\cite{Y.Wang}. 
By comparison, in the density matrix calculation, $\tau$ is introduced through the imaginary part of the retarded/advanced Green's function as $\hat G^{R/A}(\varepsilon)=\frac{1}{\varepsilon-\hat H_0\pm i\eta}$~\cite{H.Watanabe2020, H.Watanabe2021}, with the linewidth $\eta=\tau^{-1}$. Therefore, the dependence of $\gamma'$ on the impurity scatterings is expected for the quantum oscillations due to the formation of the LLs. 

Figure~\ref{Fig3}(c) plots $\gamma'$ at $\eta=1$ and $2$ meV, for which $\sigma_{xxx}$ takes relatively large values. 
We observe that (i) $\gamma'$ exhibits a decreasing trend with the magnetic field $B$, which can capture the experimental results qualitatively~\cite{Y.Wang, N.Wang};
and (ii) $\gamma'$ reaches the order of $10^{-10}$ T$^{-1}$A$^{-1}$m$^2$, and agrees with the SBE calculation $\gamma'\sim10^{-11}$ T$^{-1}$A$^{-1}$m$^2$~\cite{Y.Wang}.  Compared with the experimental value $\gamma'\sim10^{-7}$ T$^{-1}$A$^{-1}$m$^2$~\cite{Y.Wang, N.Wang}, our results still exhibit a large discrepancy, which may be attributed to the unavoidable presence of charged impurities and charged puddles~\cite{B.Skinner, N.Borgwardt}.

\section{Effect of the $\Delta$ term}

\begin{figure*}
	\centering
	\includegraphics[width=18cm]{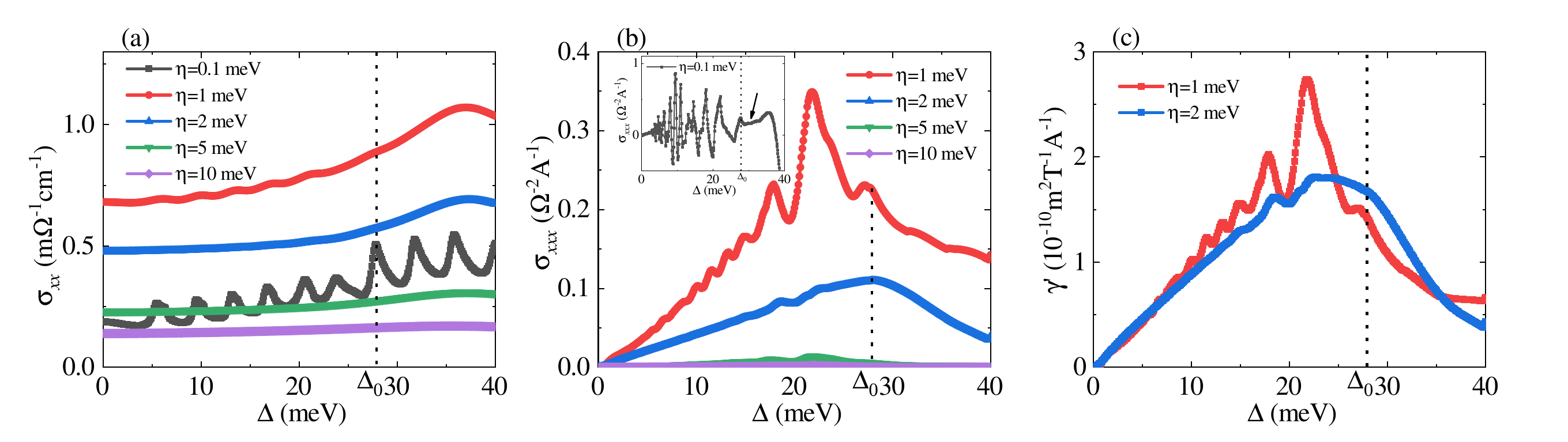}
	\caption{(Color online) (a)$-$(b) The conductivities $\sigma_{xx}$ and $\sigma_{xxx}$ versus $\Delta$ for different linewidths $\eta$.  The inset in (b) shows that with  $\eta=0.1$ meV, $\sigma_{xxx}$ exhibits strong oscillations.  When $\Delta\geq\Delta_0$, the almost unchanged $\sigma_{xxx}$ is pointed out by the arrow.  (c) The MCA coefficient $\gamma'$ versus $\Delta$ when $\eta=1$ and $2$ meV.  We take the parameter $\xi=6$ meV, the magnetic field $B=0.2$ T, and the chemical potential $\mu=30$ meV.}  
	\label{Fig4} 
\end{figure*}

Since the $xz$-mirror symmetry-breaking $\Delta$ term plays a decisive role in the selection rules, we study its influence on MCA.  With the magnetic field setting as $B=0.2$ T, in Figs.~\ref{Fig4}(a) and~\ref{Fig4}(b), the conductivities, $\sigma_{xx}$ and $\sigma_{xxx}$ are displayed as functions of $\Delta$, respectively.  In the limiting clean case of $\eta=0.1$ meV, we see that $\sigma_{xx}$ exhibits certain oscillations with $\Delta$ [Fig.~\ref{Fig4}(a)], which are caused by the $\Delta$ term-driven LL crossings over the chemical potential.  
With the linewidth $\eta\geq1$ meV, the oscillations are smeared by the impurity scatterings and $\sigma_{xx}$ increases slowly with $\Delta$.

For $\sigma_{xxx}$, when $\eta=0.1$ meV, it also exhibits strong oscillations with $\Delta$ [Fig.~\ref{Fig4}(b), inset].  Compared with $\sigma_{xx}$, more oscillations are visible in $\sigma_{xxx}$.  With increasing impurity scatterings, the neighboring peaks of $\sigma_{xxx}$ merge together.  When $\eta=1$ meV, we see that the oscillations lie on an increasing background [Fig.~\ref{Fig4}(b)]. 
If $\Delta=0$, the parity symmetry of the system is preserved and the selection rules take the conventional ones $n\rightarrow n\pm1$ ~\cite{Y.X.Wang2022, T.Devakul}, thus $\sigma_{xxx}$ vanishes.  This conclusion is consistent with the previous tilted Weyl semimetal study~\cite{S.Das}, in which the second-order longitudinal conductivity vanishes without inversion symmetry breaking.  
At a finite $\Delta$, the parity symmetry is broken and the selection rules are mixed, leading to the nonvanishing $\sigma_{xxx}$.  Since more selection rules are allowed, meaning that more electronic states can participate in the magnetotransport, thus $\sigma_{xxx}$ shows a pronounced enhancement with $\Delta$.  
When $\eta=2$ meV, $\sigma_{xxx}$ also increases linearly with $\Delta$, and the oscillations disappear. 

When $\Delta$ further increases to $\Delta\geq\Delta_0=27.8$ meV, with $\eta=0.1$ meV, we observe that (i) the peak heights of $\sigma_{xx}$ get enhanced [Fig.~\ref{Fig4}(a)], as the additional saddle points of the inverted LLs that are located at the finite wave vectors $\pm k_z$ move across the chemical potential (see Sec. III in the SM~\cite{SuppMat});  
and (ii) $\sigma_{xxx}$ becomes almost unchanged, as pointed out by the arrow [Fig.~\ref{Fig4}(b), inset].  This is because $\sigma_{xxx}$ is dominated by the electronic states of the saddle points; now around the chemical potential, the LL energies at the $\Gamma$ point become stable and the additional saddle points of the transitioned LLs have different $k_z$ (see Sec. IV in the SM~\cite{SuppMat}).  In both $\eta=1$ and $2$ meV cases, the weak impurity scatterings will drive the decreasing of $\sigma_{xxx}$ with $\Delta$ [Fig.~\ref{Fig4}(b)]. 

The MCA coefficient $\gamma'$ is determined by the combined effects of $\sigma_{xx}$ and $\sigma_{xxx}$, and the results are displayed in Fig.~\ref{Fig4}(c).  
When $\eta=1$ meV, $\gamma'$ first oscillates with $\Delta$ on a linear increasing background and then decreases, in which the crossing point occurs at $\Delta=\Delta_0$.  When $\eta=2$ meV, $\gamma'$ exhibits a similar variation, but with no oscillations.  The trend of $\gamma'$ at large $\Delta$ is consistent with the SBE calculations~\cite{Y.Wang}, where $\gamma'$ was suggested to be inversely proportional to $\Delta$ when $\Delta\gg\mu$.

\section{Discussions and Conclusions}

We discuss the effect of the Zeeman term caused by the magnetic field.  The corresponding Hamiltonian is $\hat H_{Z1}=-\frac{1}{2}g_1\mu_BB\sigma_z$ or $\hat H_{Z2}=-\frac{1}{2}g_2\mu_BB\sigma_z\otimes\tau_z$~\cite{Z.Cai, S.Galeski2022}, where  $g_{1(2)}$ denotes the Land\'e $g$ factor and $\mu_B$ is the Bohr magneton.  Although the Zeeman term can further split the $\lambda=\pm1$ LL branches, it will not break the parity property of the LLs, as the Zeeman term commutes with the parity operator $\hat{\cal P}^{-1}\hat H_{Z1(2)}\hat{\cal P}=\hat H_{Z1(2)}$.  Thus the selection rules as well as our main conclusions will remain unchanged to the Zeeman term. 

To summarize, in this paper, by developing a full quantum theory, we calculate the MCA coefficient $\gamma'$ in 3D ZrTe$_5$ and explore the effects of the magnetic field, the impurity scatterings and the $xz$-mirror symmetry breaking.  We reveal the role played by the $xz$-mirror symmetry breaking in forming the mixed selection rules, which in turn determines the nonvanishing $\sigma_{xxx}$ and $\gamma'$.  The results show that $\gamma'$ can capture the qualitative features of the experiments~\cite{Y.Wang, N.Wang}, and exhibits a nonmonotonous dependence on the strength of the $xz$-mirror symmetry breaking.  

Since the full quantum theory is based on the LL energies and wave functions, 
we expect that the derived $\sigma_{xxx}$ can be used to analyze the recently reported MCA in the Weyl semimetal WTe$_2$~\cite{T.Yokouchi}, and the formula of $\sigma_{\alpha\beta\gamma}$ can be further extended to study the nonlinear Hall effect in bilayer and few-layer WTe$_2$~\cite{Q.Ma, K.Kang}.

\section{Acknowledgments} 

This paper was supported by the Natural Science Foundation of China (Grants No. 11804122, No. 11905054, and No. 12275075), the National Key Research and Development Program of Ministry of Science and Technology (Grant No. 2021YFA1200700), the Fundamental Research Funds for the Central Universities from China, and the China Postdoctoral Science Foundation (Grant No. 2021M690970).

\end{document}